\documentclass[preprint,prd,nofootinbib,tightenlines,amsmath,amssymb]{revtex4}
\usepackage{enumerate}
\usepackage{multirow}
\usepackage[utf8]{inputenc}
\usepackage{anysize}
\usepackage{textcomp}
\usepackage{epsfig}
\usepackage{graphics,color} 
\usepackage{verbatim}
\usepackage{afterpage}
\usepackage{amsmath}    % need for subequations
\usepackage{soul}
\usepackage{xcolor,cancel}
\DeclareUnicodeCharacter{00A0}{ }

\usepackage{blindtext}

\catcode`\@=11
\def\sla@#1#2#3#4#5{{%
 \setbox\z@\hbox{$\m@th#4#5$}%
 \setbox\tw@\hbox{$\m@th#4#1$}%
 \dimen4\wd\ifdim\wd\z@<\wd\tw@\tw@\else\z@\fi
 \dimen@\ht\tw@
 \advance\dimen@-\dp\tw@ \advance\dimen@-\ht\z@
 \advance\dimen@\dp\z@
 \divide\dimen@\tw@ \advance\dimen@-#3\ht\tw@
 \advance\dimen@-#3\dp\tw@ \dimen@ii#2\wd\z@
 \raise-\dimen@\hbox to\dimen4{%
 \hss\kern\dimen@ii\box\tw@\kern-\dimen@ii\hss}%
 \llap{\hbox to\dimen4{\hss\box\z@\hss}}}}

\def\cpto{\mathrel {\vcenter {\baselineskip 0pt \kern 0pt
    \hbox{$H_{r.f.}$} \kern 0pt \hbox{$\longrightarrow$} }}}
\def\slashed#1{%
 \expandafter\ifx\csname sla@\string#1\endcsname\relax
{\mathpalette{\sla@/00}{#1}}
\fi}
\def\declareslashed#1#2#3#4#5{%
 \expandafter\def\csname sla@\string#5\endcsname{%
#1{\mathpalette{\sla@{#2}{#3}{#4}}{#5}}}}
 \catcode`\@=12
\declareslashed{}{/}{.08}{0}{D}
 \declareslashed{}{/}{.1}{0}{A}
 \declareslashed{}{/}{0}{-.05}{k}
 \declareslashed{}{/}{.1}{0}{\partial}
 \declareslashed{}{\not}{-.6}{0}{f}

\def\lsim{\mathrel {\vcenter {\baselineskip 0pt \kern 0pt
    \hbox{$<$} \kern 0pt \hbox{$\sim$} }}}
\def\gsim{\mathrel {\vcenter {\baselineskip 0pt \kern 0pt
    \hbox{$>$} \kern 0pt \hbox{$\sim$} }}}

\newcommand{\bea}{\begin{eqnarray}}
\newcommand{\eea}{\end{eqnarray}}

\begin{document}

\baselineskip=15pt
\preprint{}

\title{Top and bottom tensor couplings from a color octet scalar}

\author{Roberto Martinez$^1$\footnote{Electronic address: remartinezm@unal.edu.co},  German Valencia$^{2}$\footnote{Electronic address: German.Valencia@monash.edu }}
\affiliation{
$^{1}$Departamento de F\'{i}sica, Universidad Nacional de Colombia,
Ciudad Universitaria, K. 45 No. 26-85, Bogot\'{a} D.C., Colombia\\
$^{2}$School of Physics and Astronomy, Monash University, Wellington Road, Melbourne, Victoria 3800, Australia.\footnote{On leave from Department of Physics, Iowa State University, Ames, IA 50011.}}

\date{\today}

\vskip 1cm
\begin{abstract}
We compute the one-loop contributions from a color octet scalar to the tensor anomalous couplings of top and bottom quarks to gluons, photons and $W$ bosons. We use known constraints on the parameters of the model to compare the predicted size of these couplings with existing phenomenological constraints.
\end{abstract}

\pacs{PACS numbers: }

\maketitle

\section{Introduction}

One  of the main tasks of the LHC is to measure the couplings between quarks and gauge bosons precisely in order to search for new physics through possible deviations from their standard model (SM) values. All the possible deviations from the SM couplings have been catalogued up to dimension six with an effective Lagrangian that is consistent with the symmetries of the SM  \cite{Buchmuller:1985jz,Grzadkowski:2010es}. Of particular interest are the couplings of the top-quark because many ideas for new physics stem from the large value of its mass, and its couplings can be probed by the LHC which is the first top-quark factory. Related bottom-quark couplings may also receive large corrections in models where top-quark couplings are enhanced.

Amongst the anomalous top-quark couplings one finds the flavor diagonal dipole-type couplings to photons and gluons. These are simply the anomalous magnetic moment (MDM), the electric dipole moment (EDM), and their color generalizations CMDM and CEDM respectively. These couplings introduce spin correlations between top and anti-top pairs beyond those present in the SM, and have thus received much attention because the weak decay of the top-quark allows one to analyze its spin. They are usually parametrized as\footnote{In the notation of \cite{Martinez:2007qf} $\Delta\kappa = 4m_ta_t^g$.}
\begin{eqnarray}
{\cal L}=
\frac{e}{2}\ \bar{f}\ \sigma^{\mu\nu}\left(a_f^\gamma+i\gamma_5 d_f^\gamma\right) \ f\ F_{\mu\nu}+\frac{g_s}{2}\ \bar{f}\ T^a\sigma^{\mu\nu}\left(a_f^g+i\gamma_5 d_f^g\right) \ f\ G^a_{\mu\nu}.
\label{defcoup}
\end{eqnarray}
LHC related phenomenological studies have illustrated possible constraints on these couplings proposing a variety of observables: deviations from SM cross-sections \cite{Hioki:2012vn,Hioki:2013hva}, triple product (`T-odd') correlations \cite{Sjolin:2003ah,Gupta:2009wu,Barger:2011pu,Bernreuther:2013aga,Biswal:2012dr,Rindani:2015vya,Hayreter:2015ryk,Mileo:2016mxg}, new CP-even spin correlations \cite{Baumgart:2012ay,Bernreuther:2015yna}, lepton energy asymmetries \cite{HIOKI:2011xx}, and associated production of Higgs bosons \cite{Hayreter:2013kba}. The bottom-quark CEDM and CMDM couplings have been studied in associated production with a Higgs boson \cite{Hayreter:2013kba,Bramante:2014hua}. To study the  photon couplings, rare decays such as  $B\to X_s\gamma$, as well as $t\bar{t}\gamma$ cross-sections have been proposed \cite{Bouzas:2012av}.

A related set of anomalous couplings, the transition dipole moments $f_T^{L,R}$, occurs in the charged coupling $tbW$
\begin{eqnarray}
{\cal L}_{tbW}=\frac{g}{\sqrt{2}}\bar{b}\gamma^\mu\left(f_V^LP_L+f_V^RP_R\right)tW^-_\mu -\frac{g}{\sqrt{2}}\bar{b}\frac{\sigma^{\mu\nu}\partial_\nu W^-_\mu}{M_W}\left(f_T^LP_L+f_T^RP_R\right)t + {\rm h.c.}
\label{defch}
\end{eqnarray}
LHC related phenomenological studies for these couplings also exist, for example the study of $W$ helicity fractions and angular asymmetries \cite{Fischer:2001gp,Do:2002ky,AguilarSaavedra:2007rs} as well as T-odd observables \cite{AguilarSaavedra:2010nx,Rindani:2011gt,Sahin:2012ry}. Recent overviews of anomalous couplings in the top-quark and Higgs sectors can be found in  Ref.~\cite{Ellis:2014dva,Englert:2014uua,Englert:2014oea,Cirigliano:2016njn,Cirigliano:2016nyn}.

The size of these anomalous couplings has been considered in a variety of cases: 331 models, topcolor models and extra dimension models \cite{Martinez:2007qf}; two Higgs doublet models \cite{Martinez:2001qs,Martinez:2007qf,GonzalezSprinberg:2011kx,Duarte:2013zfa,Gaitan:2015aia,Arhrib:2016vts}; models with vector like multiplets \cite{Ibrahim:2011im}; and MSSM extensions \cite{Aboubrahim:2015zpa}. Their one-loop value within the SM has also been computed in some cases \cite{Martinez:2007qf,GonzalezSprinberg:2011kx}, and it is known that the EDMs vanish at this order within the SM.

An interesting extension of the SM results from considering an additional color octet scalar,  as introduced some time ago by Manohar and Wise (MW) \cite{Manohar:2006ga}. This particular extension of the SM is motivated by the requirement of minimal flavor violation, with which new physics naturally satisfies constraints on flavor changing neutral currents \cite{Chivukula:1987py,D'Ambrosio:2002ex}. In this paper we consider the contribution of the new scalars that appear in the MW model to the  anomalous couplings described above. Since the couplings of these scalars to quarks are proportional to the quark masses, this model is an example  where larger effects are expected for top and bottom anomalous couplings.

\section{The Model}

The MW model contains a number of parameters that have been studied phenomenologically. In particular the new color octet scalars  have a large effect on loop level Higgs production and decay \cite{Manohar:2006ga}. They are also constrained by precision measurements \cite{Gresham:2007ri,Burgess:2009wm}, one-loop effective Higgs couplings \cite{He:2011ti,Dobrescu:2011aa,Bai:2011aa,Cacciapaglia:2012wb,Cao:2013wqa,He:2013tia}, flavor physics \cite{Grinstein:2011dz,Cheng:2015lsa}, unitarity and vacuum stability \cite{Reece:2012gi,He:2013tla,Cheng:2016tlc} and other LHC processes \cite{Gerbush:2007fe,Arnold:2011ra,He:2011ws,Kribs:2012kz}.

In the MW model, the inclusion of the new field $S$ transforming as $(8,2,1/2)$ under the SM gauge group $SU(3)_C\times SU(2)_L\times U(1)_Y$ introduces several new, renormalizable, interaction terms to the Lagrangian. Because $S$ has non-trivial SM quantum numbers, it will have the corresponding gauge interactions. In addition there will be new terms in the Yukawa couplings that are consistent with minimal flavor violation \cite{Manohar:2006ga}, 
\begin{equation}
{\cal L}_Y=-\eta_U e^{i\alpha_{U}} g^{U}_{ij}\bar{u}_{Ri}T^a Q_j S^a - \eta_D e^{i\alpha_{D}}g^{D}_{ij}\bar{d}_{Ri}T^a Q_j S^{\dagger a} + h.c,
\end{equation}
where $Q_i$ are left-handed quark doublets, $S=S^aT^a \;(a=1,...,8)$ and the $SU(3)$ generators are normalized as ${\rm Tr} (T^aT^b)=\delta^{ab}/2$. The  matrices $g^{U,D}_{ij}$ are the same as the quark Yukawa couplings, and $\eta_{U,D}$ along with their phases $e^{i\alpha_{U,D}}$, are new overall factors (that can be complex and we write the phases explicitly). Non-zero phases would signal $CP$ violation beyond the SM and contribute to the EDM and CEDM of quarks. In the quark mass eigenstate basis these couplings are given by
\begin{eqnarray}
{\cal L}_Y &=& - \frac{\sqrt{2}}{v}\eta_U e^{i\alpha_{U}}  \bar U_R T^a \hat M^u U_LS^{a0} + \frac{\sqrt{2}}{v} \eta_U e^{i\alpha_{U}}  \bar U_R T^a \hat M^u V_{KM} D_L S^{a+}\nonumber\\
&-& \frac{\sqrt{2}}{ v} \eta_D e^{i\alpha_{D}}\bar D_R T^a \hat M^d D_LS^{a0\dagger} - \frac{\sqrt{2}}{ v}\eta_D e^{i\alpha_{D}} \bar D_R T^a \hat M^u V^\dagger_{KM} U_L S^{a-} + {\rm ~h.c.},
\end{eqnarray}
where $\hat{M}^{u,d}$ are the diagonal quark mass matrices, $\hat{M}^{u,d} = {\rm diag}(m_{u,d}, m_{c,s}, m_{t,b})$; the quark fields are 
$U_{L,R} = {\rm diag}(u_{L,R}, c_{L,R}, t_{L,R})$ and $D_{L,R} = {\rm
diag}(d_{L,R}, s_{L,R}, b_{L,B})$. The neutral complex field $S^{a0}$ can be further decomposed into a scalar $S^{a0}_R$ and a pseudo-scalar $S^{a0}_I$ as $S^{a0} = (S^{a0}_R + i\ S^{a0}_I)/\sqrt{2}$.  The following combinations of constants  usually appear together, and in our results we will use the shorthand:
\begin{equation}
g_t \equiv \left(\eta_U \frac{m_t}{v}\right),\ \ g_b\equiv\left(\eta_D \frac{m_b}{v}\right).
\end{equation}

The most general renormalizable scalar potential for this model is given in Ref.~\cite{Manohar:2006ga} and contains many terms. Of these, our calculation will only depend on the terms contributing to the  mass of the new scalars,
\begin{eqnarray}
V&=&\lambda\left(H^{\dagger i}H_i-\frac{v^2}{2}\right)^2+2m_s^2\ {\rm Tr}S^{\dagger i}S_i +\lambda_1\ H^{\dagger i}H_i\  {\rm Tr}S^{\dagger j}S_j +\lambda_2\ H^{\dagger i}H_j\  {\rm Tr}S^{\dagger j}S_i 
\nonumber \\
&+&\left( \lambda_3\ H^{\dagger i}H^{\dagger j}\  {\rm Tr}S_ iS_j .
+{\rm ~h.c.}\right)
\label{potential}
\end{eqnarray}
Here $v\sim 246$~GeV is the Higgs
vacuum expectation value (vev) with $\langle H \rangle = v/\sqrt{2}$, 
the traces are over the color indices and the $SU(2)$ indices $i,j$ are displayed explicitly. Furthermore, $\lambda_3$ can be chosen to be real by a suitable definition of $S$. After symmetry breaking, the non-zero vev of the Higgs gives the physical Higgs scalar $h$  a mass $m^2_H = 2 \lambda v^2$ and it also splits  the octet scalar masses as,
\begin{eqnarray}
m^2_{S^{\pm}} =  m^2_S + \lambda_1 \frac{v^2}{4},&&
m^2_{S^{0}_{R,I}} =  m^2_S + \left(\lambda_1 + \lambda_2 \pm 2 \lambda_3 \right) \frac{v^2}{4},
\end{eqnarray}
The parameters $m_S^2$, and $\lambda_{1,2,3}$ should be chosen such that the above squared masses remain positive. Relations between various parameters follow from custodial symmetry,  and we will use $2\lambda_3=\lambda_2$ which makes $S_I$ and $S^\pm$ degenerate.

\section{Anomalous couplings}

\subsection{Top CMDM and CEDM}

\begin{figure}[thb]
\includegraphics[width=.65\textwidth]{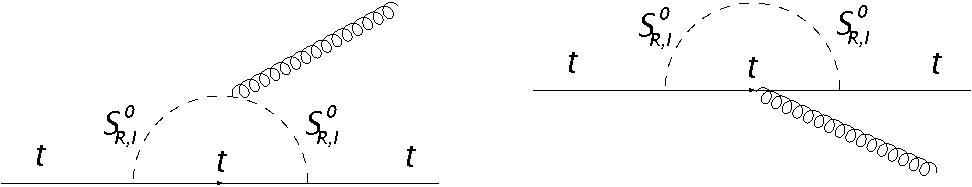}
\caption{Diagrams with neutral scalars contributing to $a^g_{t,b}$ and $d^g_{t,b}$ as described in the text.
\label{feyn-dia}}
\end{figure}

We begin with the dipole-type couplings of the top-quark  to the gluon, the chromo magnetic dipole moment $a^g_{t}$ and chromo electric dipole moment  $d^g_{t}$ in the notation of Eq.~\ref{defcoup}. In the one-loop calculation these couplings are finite since they do not appear in the tree-level Lagrangian and receive several contributions. 
One such contribution arises from loops involving neutral scalars as depicted in the two diagrams shown in Figure~\ref{feyn-dia}.  
Denoting by $M_S$ the mass of the corresponding scalar in the loop and using $r_{tR}=m_t/M_{S_R}$, $r_{tI}=m_t/M_{S_I}$, these two diagrams  result in
\begin{eqnarray}
a^g_t &=& \frac{m_tg_t^2}{16\pi^2 M^2_{S_R}}  \left(\frac{3}{2}(F_{2,1}(r_{tR})-2c_U^2F_{1,1}(r_{tR}))-\frac{1}{6}(F_{3,0}(r_{tR})-2c_U^2F_{2,0}(r_{tR}))\right) \nonumber \\
&+& \frac{m_tg_t^2}{16\pi^2 M^2_{S_I}}  \left(\frac{3}{2}(F_{2,1}(r_{tI})-2s_U^2F_{1,1}(r_{tI}))-\frac{1}{6}(F_{3,0}(r_{tI})-2s_U^2F_{2,0}(r_{tI}))\right) \nonumber \\
d^g_t &=& \frac{s_Uc_Um_t}{8\pi^2} g_t^2 \left[\frac{\left(\frac{3}{2}F_{1,1}(r_{tR})-\frac{1}{6}F_{0,0}(r_{tR})\right)}{M_{S_R}^2}-\frac{\left(\frac{3}{2}F_{1,1}(r_{tI})-\frac{1}{6}F_{0,0}(r_{tI})\right)}{M_{S_I}^2}\right].
\label{topcedm}
\end{eqnarray}
The factors $3/2$ and $-1/6$ are the color factors for the diagrams on the left and right respectively, and the form factors $F_{i,j}(r_t)$ are one parameter Feynman integrals explicitly given in the Appendix. We have also introduced the short hand notation $c_U=\cos\alpha_U$, $s_U=\sin\alpha_U$.

\begin{figure}[thb]
\includegraphics[width=.65\textwidth]{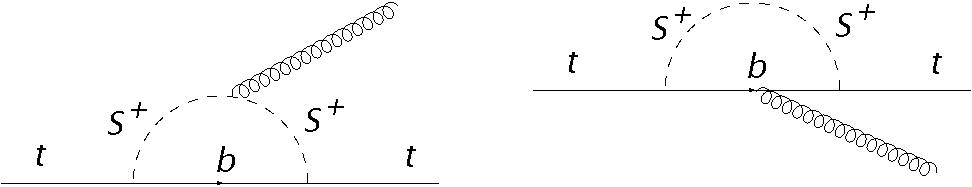}
\caption{Diagrams with charged scalars contributing to $a^g_{t,b}$ and $d^g_{t,b}$ as described in the text.
\label{feyn-dia2}}
\end{figure}

There are also two diagrams with charged scalars and a bottom-quark in the loop as shown in Figure~\ref{feyn-dia2}, and they result in 
\begin{eqnarray}
a^g_t &=&   -\frac{m_t |V_{tb}|^2}{16\pi^2M_{S^+}^2}\left[\left(g_t^2+g_b^2\right)\left(\frac{3}{2}G_{1,2}(r_t,r_b)-\frac{1}{6}G_{2,1}(r_t,r_b)\right)\right. \nonumber \\
&+&\left. 2\frac{m_b}{m_t}g_tg_b \cos(\alpha_D+\alpha_U)\left(\frac{3}{2}G_{1,1}(r_t,r_b)+\frac{1}{6}G_{2,0}(r_t,r_b)\right)\right] \nonumber \\
d_t^g &=&-  \frac{m_t|V_{tb}|^2}{16\pi^2M_{S^+}^2} 2\frac{m_b}{m_t}g_tg_b\sin(\alpha_D+\alpha_U) \left(\frac{3}{2}G_{1,1}(r_t,r_b)-\frac{1}{6}G_{2,0}(r_t,r_b)\right) 
\label{tcharged}
\end{eqnarray}
where the factors $3/2$ and $-1/6$ are again the color factors for the diagrams on the left and right respectively and $G_{i,j}(r_1,r_2)$ are one parameter Feynman integrals also shown in the Appendix. In the limit $m_b\to0$, Eq.~\ref{tcharged} simplifies considerably, with $d_t^g$ vanishing and $a_t^g$ retaining a term proportional to the top-quark mass (cubed),
\begin{eqnarray}
a^g_t &\approx&  - \frac{m_t |V_{tb}|^2}{16\pi^2M_{S^+}^2}g_t^2\left(\frac{3}{2}G_{1,2}(r_t,0)-\frac{1}{6}G_{2,1}(r_t,0)\right).
\end{eqnarray}

\begin{figure}[thb]
\includegraphics[width=3.5in]{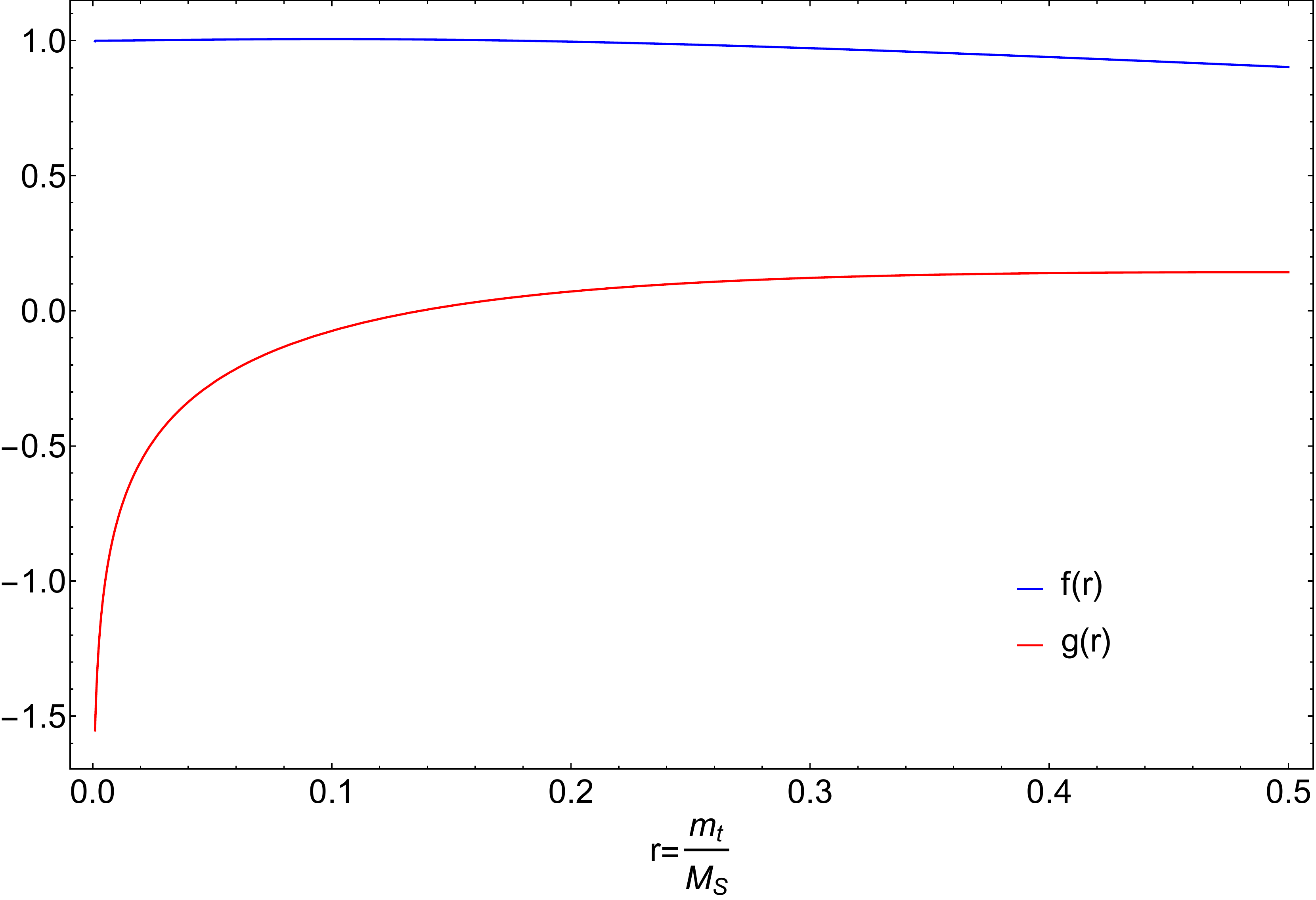}
\caption{Numerical evaluation of the functions $f(r)$ and $g(r)$ appearing in Eq.~\ref{totres}.
\label{numfun}}
\end{figure}

To combine the diagrams to obtain the final result it is important to note that the two neutral resonances tend to cancel each other's contribution to CP violation \cite{He:2011ws}, and in fact produce a vanishing CEDM when they have the same mass. To extract the leading behaviour in the mass difference, we write the scalar masses in the custodial symmetry limit as 
\begin{eqnarray}
M_{S_R}^2=M_{S}^2+\frac{v^2}{2}\lambda_2, && M_{S_I}^2=M_{S^{\pm} }^2\equiv M_S^2
\end{eqnarray}
and work to leading order in $\lambda_2$.  We can then add all the contributions neglecting $m_b$ and setting $V_{tb}=1$ to finally obtain 
\begin{eqnarray}
a^g_t &=&-\frac{7}{12} \frac{m_t\eta_U^2}{16\pi^2v^2}  r_t^2 \ f(r_t),\nonumber \\
 d^g_t &=& -\frac{m_t\eta_U^2}{16\pi^2  M_S^2} \lambda_2s_Uc_U\ r_t^2 \ g(r_t),
 \label{totres}
\end{eqnarray}
where the functions $f(r),g(r)$ are combinations of one dimensional Feynman parameter integrals and we show their numerical value in Figure~\ref{numfun}.

\subsection{Bottom CMDM and CEDM}

Similar diagrams, with the roles of top and bottom-quarks interchanged, generate $a_b^g$ and $d_b^g$. The two diagrams with neutral scalars in the loop also have a bottom-quark in the loop, and are proportional to $m_b$ (cubed). Defining $r_{bR,I}=m_b/M_{S_{R,I}}$ these two diagrams give 
\begin{eqnarray}
a^g_b &=& \frac{m_b}{16\pi^2M_{S_R}^2} g_b^2 \left(\frac{3}{2}(F_{2,1}(r_{bR})-2c_D^2F_{1,1}(r_{bR}))-\frac{1}{6}(F_{3,0}(r_{bR})-2c_D^2F_{2,0}(r_{bR}))\right)\nonumber \\
&+&\frac{m_b}{16\pi^2M_{S_I}^2}g_b^2 \left(\frac{3}{2}(F_{2,1}(r_{bI})-2s_D^2F_{1,1}(r_{bI}))-\frac{1}{6}(F_{3,0}(r_{bI})-2s_D^2F_{2,0}(r_{bI}))\right) \nonumber \\
&\approx& -\frac{7}{18}\  \frac{\eta_D^2 m_b}{16\pi^2 v^2}r_b^2 \nonumber \\
d^g_b &=& \frac{s_Dc_Dm_b}{8\pi^2} g_b^2\left(\frac{ \left(\frac{3}{2}F_{1,1}(r_{bR})-\frac{1}{6}F_{0,0}(r_{bR})\right)}{M_{S_R}^2}-\frac{ \left(\frac{3}{2}F_{1,1}(r_{bI})-\frac{1}{6}F_{0,0}(r_{bI})\right)}{M_{S_I}^2}\right)\nonumber \\
&\approx &  -\frac{\eta_D^2 m_b}{16\pi^2M_{S}^2}\ c_Ds_D\lambda_2 r_b^2\left(\frac{3}{4}+\log\frac{m_b}{3M_S}\right),
\label{bcedm}
\end{eqnarray}
where we have used the short hand notation $c_D=\cos\alpha_D$, $s_D=\sin\alpha_D$.

The two diagrams with charged scalars now have a top-quark in the loop and are analogous to Figure~\ref{feyn-dia2} interchanging $t\leftrightarrow b$ and their corresponding couplings. The result can also be obtained with this substitution from Eq.~\ref{tcharged}, and in particular it contains terms enhanced by $m_t^2/m_b^2$ with respect to Eq.~\ref{bcedm}.
\begin{eqnarray}
a^g_b &=&   -\frac{m_b |V_{tb}|^2}{16\pi^2M_{S^+}^2}\left[\left(g_t^2+g_b^2\right)\left(\frac{3}{2}G_{1,2}(r_b,r_{t})-\frac{1}{6}G_{2,1}(r_b,r_{t})\right)\right. \nonumber \\
&+&\left. 2\frac{m_t}{m_b}g_tg_b \cos(\alpha_D+\alpha_U)\left(\frac{3}{2}G_{1,1}(r_b,r_t)+\frac{1}{6}G_{2,0}(r_b,r_t)\right)\right] \nonumber \\
d_b^g &=&  -\frac{m_t|V_{tb}|^2}{16\pi^2M_{S^+}^2} 2g_tg_b\sin(\alpha_D+\alpha_U) \left(\frac{3}{2}G_{1,1}(r_b,r_t)-\frac{1}{6}G_{2,0}(r_b,r_t)\right) 
\label{bcedmc}
\end{eqnarray}

\begin{figure}[thb]
\includegraphics[width=3.5in]{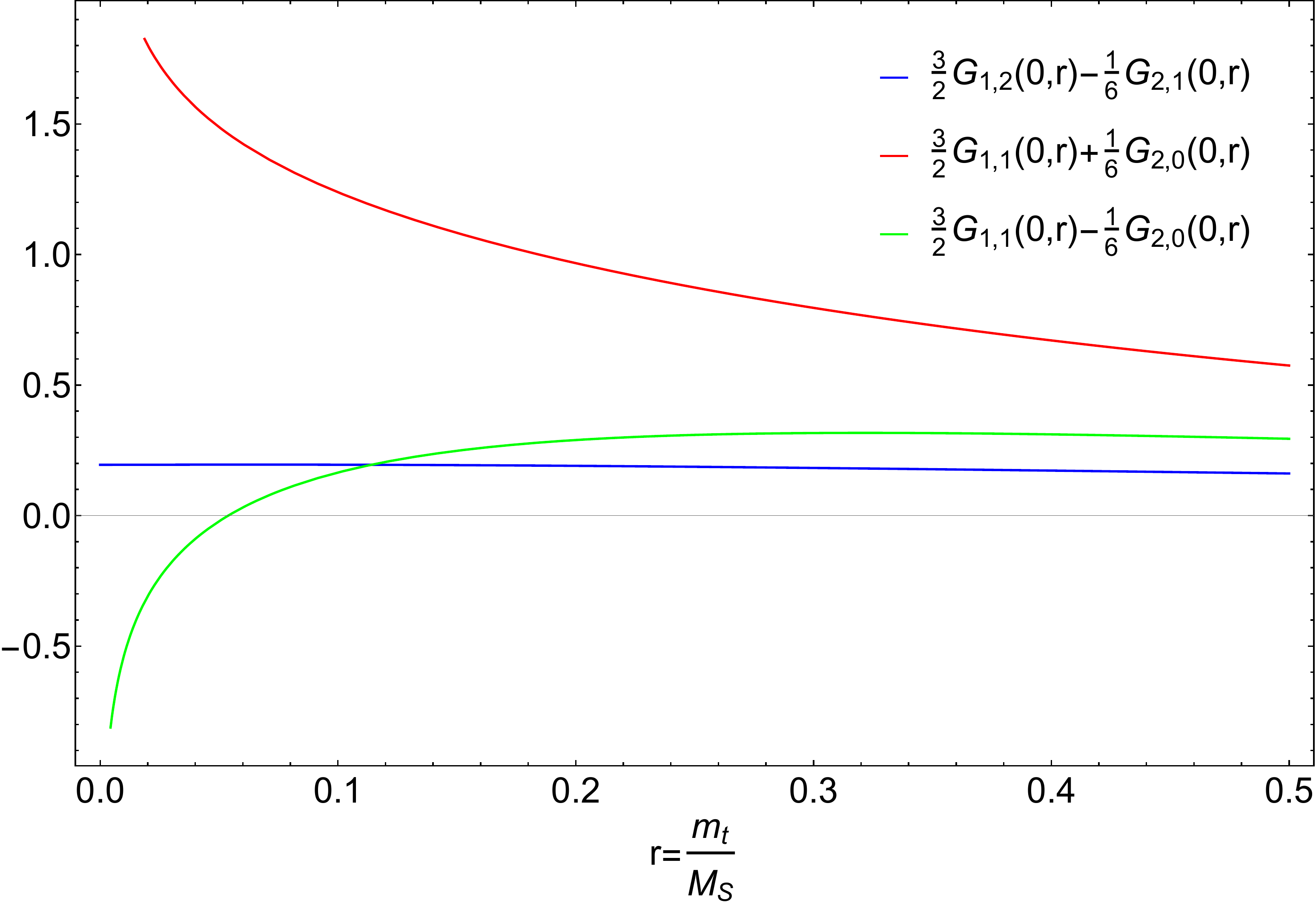}
\caption{Numerical evaluation of the functions appearing in Eq.~\ref{ceqsf}.
\label{f:cmdm-b}}
\end{figure}

For regions of parameter space where $\eta_Dm_b <<\eta_U m_t$ we can ignore the contributions from neutral scalars in the loops and the complete result simplifies to
\begin{eqnarray}
a^g_b &=& -  \frac{m_b |V_{tb}|^2}{16\pi^2v^2} r_t^2 \left[\eta_U^2\left(\frac{3}{2}G_{1,2}(0,r_t)-\frac{1}{6}G_{2,1}(0,r_t)\right)\right. \nonumber \\
&+&\left. 2\eta_U\eta_D \cos(\alpha_D+\alpha_U)\left(\frac{3}{2}G_{1,1}(0,r_t)+\frac{1}{6}G_{2,0}(0,r_t)\right)\right] \nonumber \\
d_b^g &=&  -\frac{m_b|V_{tb}|^2}{8\pi^2v^2} r_t^2 \eta_U\eta_D\sin(\alpha_D+\alpha_U) \left(\frac{3}{2}G_{1,1}(0,r_t)-\frac{1}{6}G_{2,0}(0,r_t)\right) 
\label{ceqsf}
\end{eqnarray}
The loop factors appearing in this result are evaluated numerically and plotted in Figure~\ref{f:cmdm-b}. The second term in $a^g_b$ dominates as long as $\eta_D/\eta_U\cos(\alpha_D+\alpha_U) \gsim 1/4$.

\subsection{Anomalous magnetic and  electric dipole moments}

\begin{figure}[thb]
\includegraphics[width=3.5in]{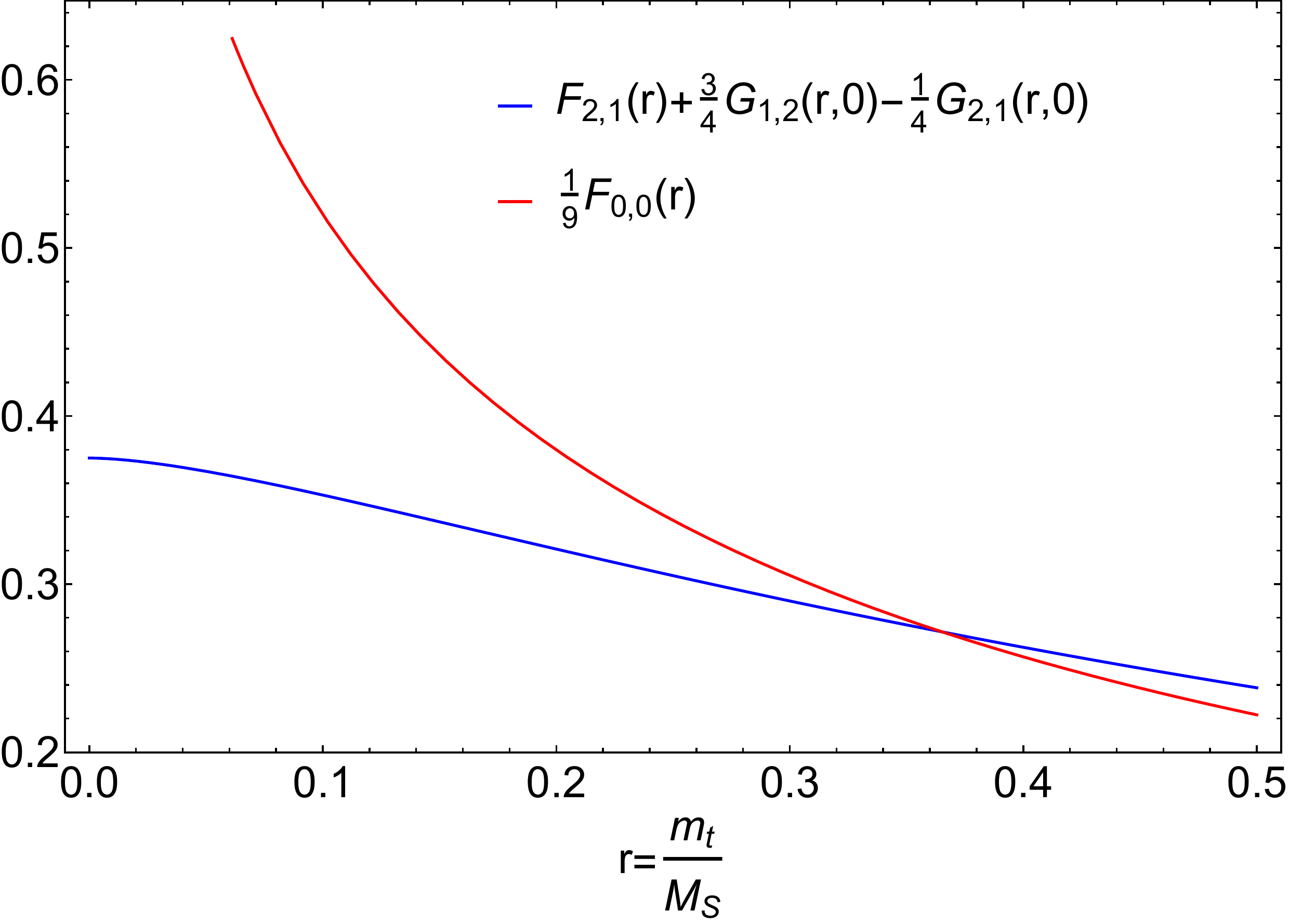}
\caption{Numerical evaluation of the functions appearing in Eq.~\ref{edmt}.
\label{f:edm-t}}
\end{figure}

Once again these couplings are finite because they do not appear in the tree level Lagrangian. They can be extracted from our previous results by interpreting the wavy lines in the Feynman diagrams  as photons: we only need to replace the color factor with the electric charge factor in the respective vertex. For neutral scalars and top-quarks in the loop the photon can only be emitted by the top-quark (diagram on the right in Figure~\ref{feyn-dia}) and the result can be obtained from Eq.~\ref{topcedm} by replacing the color factor $-1/6\ g_s$ with the new color-electric charge factor $8/9\ e$. For the diagrams involving charged scalars in Figure~\ref{feyn-dia2} we replace the color factors $3/2\ g_s$ and $-1/6\ g_s$ with $4/3\ e$ and $-4/9\ e$ respectively in Eq.~\ref{tcharged}. We obtain
\begin{eqnarray}
a^\gamma_t &=& \frac{m_tg_t^2}{16\pi^2 M^2_{S_R}}  \left(\frac{8}{9}(F_{3,0}(r_{tR})-2c_U^2F_{2,0}(r_{tR}))\right) 
+\frac{m_tg_t^2}{16\pi^2 M^2_{S_I}}  \left(\frac{8}{9}(F_{3,0}(r_{tI})-2s_U^2F_{2,0}(r_{tI}))\right) \nonumber \\
&-&   \frac{m_t |V_{tb}|^2}{16\pi^2M_{S^+}^2}\left[\left(g_t^2+g_b^2\right)\left(\frac{4}{3}G_{1,2}(r_t,r_b)-\frac{4}{9}G_{2,1}(r_t,r_b)\right)\right. \nonumber \\
&+&\left. 2\frac{m_b}{m_t}g_tg_b \cos(\alpha_D+\alpha_U)\left(\frac{4}{3}G_{1,1}(r_t,r_b)+\frac{4}{9}G_{2,0}(r_t,r_b)\right)\right] \nonumber \\
d^\gamma_t &=& \frac{s_Uc_Um_t}{8\pi^2} g_t^2 \frac{8}{9}\left(\frac{F_{0,0}(r_{tR})}{M_{S_R}^2}-\frac{F_{0,0}(r_{tI})}{M_{S_I}^2}\right) \nonumber \\
&-&  \frac{m_t|V_{tb}|^2}{16\pi^2M_{S^+}^2} 2\frac{m_b}{m_t}g_tg_b\sin(\alpha_D+\alpha_U) \left(\frac{4}{3}G_{1,1}(r_t,r_b)-\frac{4}{9}G_{2,0}(r_t,r_b)\right) 
\end{eqnarray}
To leading order in $m_b$ and $\lambda_2 v^2/M_S^2$ and setting $V_{tb}=1$, these simplify to
\begin{eqnarray}
a^\gamma_t &=& -\frac{\eta_U^2m_t}{9\pi^2 v^2}  r_t^2\left(F_{2,1}(r_t)+\frac{3}{4}G_{1,2}(r_t,0)-\frac{1}{4}G_{2,1}(r_t,0) \right)  \nonumber \\
d^\gamma_t &=&- \frac{\eta_U^2s_Uc_Um_t}{18\pi^2 M_S^2} \lambda_2 r_t^2 F_{0,0}(r_t)
\label{edmt}
\end{eqnarray}
The loop factors appearing in this result are evaluated numerically and plotted in Figure~\ref{f:edm-t}, with $V_{tb}=1$, and  $F_{0,0}$ scaled by a factor of 9 for convenience. 

\begin{figure}[thb]
\includegraphics[width=3.5in]{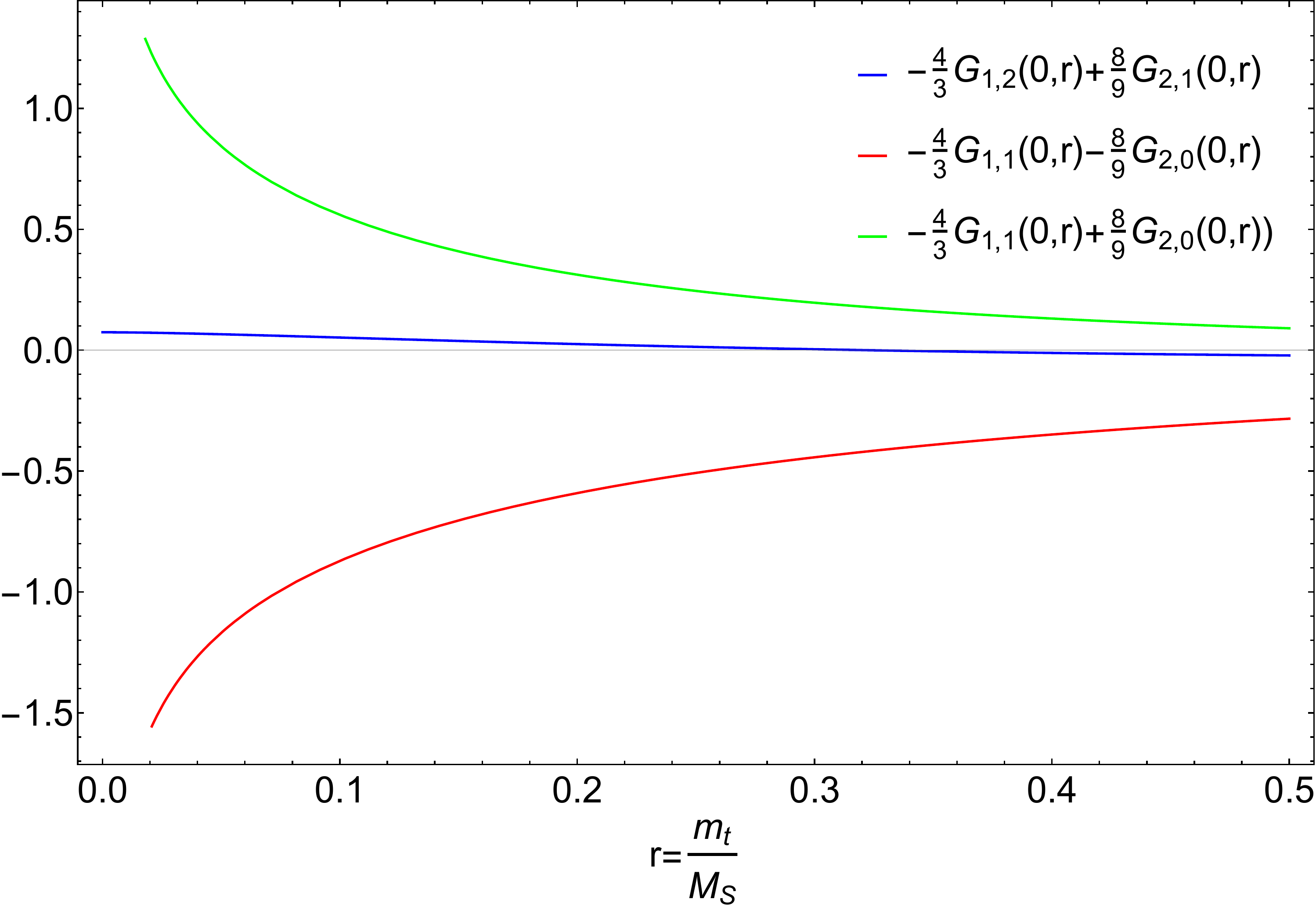}
\caption{Numerical evaluation of the functions appearing in Eq.~\ref{edmb}.
\label{f:edm-b}}
\end{figure}

Similarly, for the MDM and EDM of the bottom quark, only the diagram to the right in Figure~\ref{feyn-dia} contributes and the result can be obtained by replacing $-1/6\ g_s$ with  $-4/9\ e$ in Eq.~\ref{bcedm}. The resulting contributions are proportional to $m_b^3$ and negligible compared to those due to the exchange of a charged scalar. For the latter case the result follows from Eq.~\ref{bcedmc} replacing the color factors $3/2\ g_s$ and $-1/6\  g_s$ with $-4/3\ e$ and $8/9\ e$ respectively. Keeping only the leading terms in $m_b$ we find,
\begin{eqnarray}
a^\gamma_b &=&-   \frac{m_b |V_{tb}|^2}{16\pi^2v^2} r_t^2 \left[ \eta_U^2\left(-\frac{4}{3}G_{1,2}(r_b,r_t)+\frac{8}{9}G_{2,1}(r_b,r_t)\right)\right. \nonumber \\
&+&\left. 2\eta_U\eta_D\cos(\alpha_D+\alpha_U)\left(-\frac{4}{3}G_{1,1}(r_b,r_t)-\frac{8}{9}G_{2,0}(r_b,r_t)\right)\right] \nonumber \\
d^\gamma_b&=& - \frac{m_b|V_{tb}|^2}{8\pi^2v^2} r_t^2\eta_U\eta_D\sin(\alpha_D+\alpha_U) \left(-\frac{4}{3}G_{1,1}(r_b,r_t)+\frac{8}{9}G_{2,0}(r_b,r_t)\right) .
\label{edmb}
\end{eqnarray}
The loop factors appearing in this result are evaluated numerically and plotted in Figure~\ref{f:edm-b}. From the figure we see that $a_b^\gamma$ is dominated by the second term in the expression given in Eq.~\ref{edmb} as long as $\eta_D/\eta_U\cos(\alpha_D+\alpha_U)\gsim 1/4$.

\subsection{$tbW$ couplings}

\begin{figure}[thb]
\includegraphics[width=.7\textwidth]{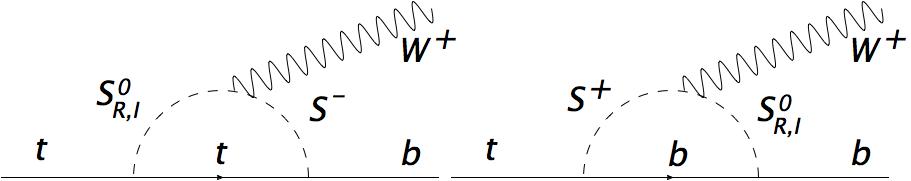}
\caption{Generic Feynman diagram contributing to $f_T^R$ as described in the text.
\label{fd2}}
\end{figure}

\begin{figure}[thb]
\includegraphics[width=3.5in]{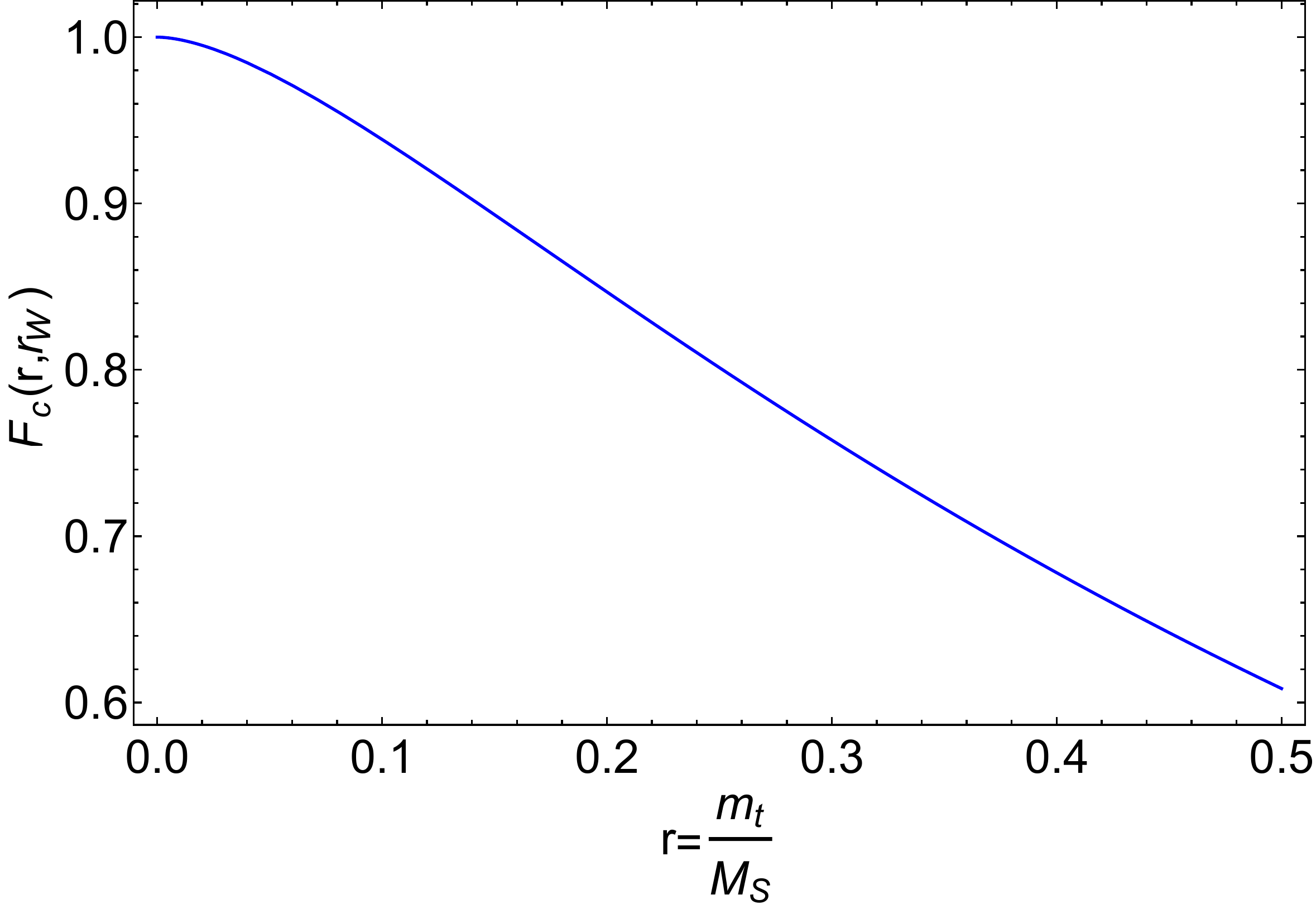}
\caption{Numerical evaluation of the function $F_c(r,r_W)$ appearing in Eq.~\ref{frtrw} for a fixed value of $r_W=m_W/m_t$.}
\label{ffc}
\end{figure}

The MW model contributes to all the couplings that appear in Eq.~\ref{defch}, but we concentrate of the tensor couplings which are finite. The contribution to $f_V^R$ is also finite and very interesting, but in this model it is suppressed by factors of $m_b$. We begin with the Feynman diagrams shown in Figure~\ref{fd2}. 
The diagrams to the left, one with $S^0_R$ and one with $S^0_I$, combine to yield the finite result (in this case the color factor is $4/3$),
\begin{eqnarray}
f_T^R=-\frac{1}{12\pi^2 }\left(\eta_U\frac{m_t}{v}\right)^2V_{tb} r_t^2 r_We^{-2i\alpha_U}F_c(r_t,r_W)
\label{leadftr}
\end{eqnarray}
where we have defined $r_t=m_t/M_S$ as before and $r_W=m_W/m_t$. The loop factor is written in the Appendix in Eq.~\ref{frtrw}, it satisfies  $F_c(0,r_W)=1$ and the $r_t$ dependence is evaluated numerically and shown in Figure~\ref{ffc}. The diagrams also induce a non-zero $f_T^L$ but it is proportional to $m_b$ and we have dropped these terms for simplicity.
Interestingly, these two diagrams also  contribute to the form factor $f_V^L$ (without an $m_b$ suppression) as inferred from the Gordon decomposition of the result  (proportional to $(p_t+p_b)^\mu\bar{b}P_R\ t$).  The form factor $f_V^L$, however, receives other contributions from these two diagrams and  from 
additional diagrams, some of which are  divergent. It is necessary to perform a full one-loop calculation, including  renormalization, to obtain a finite result for $f_V^L$. 

Diagrams to the right in Figure~\ref{fd2}, with the charged and neutral scalars interchanged are suppressed by at least $m_b^2/m_t^2$ with respect to Eq.~\ref{leadftr} and are therefore negligible except for scenarios with $\eta_D>>\eta_U$. Similarly, diagrams in which the $W$ boson is emitted from the quark line contain an $S^0_{R,I}b\bar{b}$ coupling and are also suppressed by $m_b^2/m_t^2$ with respect to Eq.~\ref{leadftr}. 

\section{Summary of results and comparison with existing constraints}

There exist a number of phenomenological papers investigating possible constraints on these anomalous couplings. We collect some of these results in Table~\ref{t:results}.

\begin{table*}[htb]
\caption{Summary of phenomenological constraints from different processes. LHC cross-sections assume these can be measured to within the $1\sigma$ uncertainty of the NLO SM calculations. LHC asymmetries are simulations at LHC14 with 10~fb$^{-1}$. Bounds quoted are $1~\sigma$ unless otherwise noted.}
\begin{center}
\begin{tabular}{|c|c|}
\hline
 Process &  Constraint \\ \hline
$\sigma(t\bar{t})$ \cite{Hayreter:2013kba}& $-0.029\lsim m_ta_t^g\lsim 0.024$ ,\  $|m_t d_t^g|\lsim 0.1$ \\  \hline
$(t\bar{t})$ T-odd \cite{Hayreter:2013kba} & $|m_t d_t^g|\lsim 0.009$  \\  \hline
$\sigma(t\bar{t}h)$ \cite{Hayreter:2013kba}& $-0.016\lsim m_ta_t^g\lsim 0.008$ ,\  $|m_t d_t^g|\lsim 0.02$ \\  \hline
neutron edm \cite{Kamenik:2011dk} & $m_td_t^g < 2 \times 10^{-3}$ (indirect) \\  \hline
$\sigma(b\bar{b}h)$\cite{Hayreter:2013kba}& $-1.3 \times 10^{-4}\lsim m_ba_b^g\lsim 2.4\times 10^{-4}$  \\ 
& $|m_b d_b^g|\lsim 1.7 \times 10^{-4}$ \\  \hline
$B\to X_s\gamma $ \cite{Grzadkowski:2008mf}& $-0.15< f_T^R < 0.57$~ (95\%C.L. )\\  \hline
 $t\to bW$ helicity fractions  \cite{AguilarSaavedra:2010nx} & $-0.0260< f_T^R < 0.0312$~ $(2\sigma)$, $f_T^L=f_V^R=0$  \\  \hline
$t\to bW$ T-odd \cite{AguilarSaavedra:2010nx}& $|$Im$(f_T^R)| \leq 0.115$ ~ $(3\sigma)$, $f_T^L=f_V^R=0$ \\  \hline
$t\to bW$ asymmetries \cite{AguilarSaavedra:2010nx}& $|$Re$(f_T^R)| \leq 0.056$ ~ $(3\sigma)$, $f_T^L=f_V^R=0$   \\  \hline
$B\to X_s\gamma $ and $\sigma(t\bar{t}\gamma)$ \cite{Bouzas:2012av} & $-1\leq m_t a_t^\gamma \leq 0.15$,  $-0.25 \leq m_t d_t^\gamma \leq 0.75$\\  \hline
$R_b$ \cite{Martinez:1997ww} & $-0.2 \leq m_t a_t^\gamma \leq 0.33$ \\  \hline
8 parameter fit & \\ 
to LHC data \cite{Hioki:2015env,Hioki:2016xtc} & $|$Re$(f_T^R)| \leq 0.120$, $|$Im$(f_T^R)| \leq 0.120$  \\  \hline
\end{tabular}
\end{center}
\label{t:results}
\end{table*}

In addition,  the couplings in the $tbW$ vertex have been recently constrained experimentally, with the 95\% C.L.  results from CMS being \cite{Khachatryan:2016sib},
\begin{eqnarray}
 |f_V^L|>0.98,\,|f_V^R|<0.16 \nonumber \\
 |f_T^L|<0.057,\, -0.049<f^R_T < 0.048
\label{t:exp}
\end{eqnarray}
A search for CP violation has also been started by CMS but they have not yet constrained $d_t^g$ \cite{CMS:2016tnr} beyond saying it is consistent with zero.

Next we compare our results to the constraints listed below in Table~\ref{t:pre}. For this purpose we have fixed $M_S$ to 500~GeV, the parameters $\eta_{U,D} =3$ and $\lambda_2 =6$. The choice for mass is just a benchmark as there are no good limits on the masses of  these type of  scalars at the LHC \cite{Cheng:2016tlc}. The value for $\eta_U$ is about tree times larger than the $1\sigma$ bound from $R_b$ \cite{Gresham:2007ri}, and nearly 50\% of its tree-level unitarity constraint \cite{He:2013tla}. Similarly, the benchmark chosen for $\lambda_2$ is well below its unitarity constraint  \cite{He:2013tla} and at the level where it saturates the parameter $S$ \cite{Manohar:2006ga}.
\begin{table*}[htb]
\caption{Summary of results.}
\begin{center}
\begin{tabular}{|c|c|}
\hline
$m_ta_t^g$ &  $-2\times10^{-3}\left|\frac{\eta_U}{3}\right|^2 \left(\frac{500~{\rm GeV}}{M_S}\right)^2$  \\ \hline
$m_td_t^g$ &  $-8 \times 10^{-4} \left|\frac{\eta_U}{3}\right|^2 \left(\frac{500~{\rm GeV}}{M_S}\right)^2 \left(\frac{\lambda_2}{6}\right) \cos\alpha_U\sin\alpha_U$ \\ \hline
$m_ba_b^g$ & $-4.4\times 10^{-6} \left|\frac{\eta_{U,D}}{3}\right|^2 \left(\frac{500~{\rm GeV}}{M_S}\right)^2 $ \\ \hline
$m_bd_b^g$ &   $-1.2 \times 10^{-6} \left|\frac{\eta_{U,D}}{3}\right|^2 \left(\frac{500~{\rm GeV}}{M_S}\right)^2\sin(\alpha_D+\alpha_U) $ \\ \hline
$m_ta_t^\gamma$ & $-2\times10^{-3}\left|\frac{\eta_U}{3}\right|^2 \left(\frac{500~{\rm GeV}}{M_S}\right)^2$  \\ \hline
$m_td_t^\gamma$ &   $-1.4\times10^{-2}\left|\frac{\eta_U}{3}\right|^2 \left(\frac{500~{\rm GeV}}{M_S}\right)^2\left(\frac{\lambda_2}{6}\right) \cos\alpha_U\sin\alpha_U$  \\ \hline
$m_ba_b^\gamma$ &  $1.8\times10^{-6} \left|\frac{\eta_{U,D}}{3}\right|^2 \left(\frac{500~{\rm GeV}}{M_S}\right)^2 $ \\ \hline
$m_bd_b^\gamma$ &  $-1.2 \times 10^{-6} \left|\frac{\eta_{U,D}}{3}\right|^2 \left(\frac{500~{\rm GeV}}{M_S}\right)^2\sin(\alpha_D+\alpha_U) $  \\ \hline
$f_T^R$ &  $-2 \times 10^{-3} \left|\frac{\eta_{U}}{3}\right|^2 \left(\frac{500~{\rm GeV}}{M_S}\right)^2 e^{-2i\alpha_U} $ \\ \hline
\end{tabular}
\end{center}
\label{t:pre}
\end{table*}
For example, keeping the values of  $\eta_{U,D}$ and $\lambda_2$ fixed, the results in Table~\ref{t:pre} can be used to interpret the bounds of Table~\ref{t:results} as constraints on $M_S$. The best LHC limit would be obtained from $\sigma(t\bar{t}h)$, $M_S \gsim 250$~GeV. Similarly the best overall limit, from the neutron edm, implies $M_S\gsim 316$~GeV. Both of these limits are slightly better than the existing robust LEP limit $M_S \gsim 100$~GeV \cite{Burgess:2009wm}. A rough estimate for the production cross-section of a single $S^0_{R,I}$ with these masses can be found in \cite{Gresham:2007ri,Burgess:2009wm,He:2011ws}, however, this cross-section depends strongly on the parameters $\lambda_{4,5}$ which do not affect this work. Setting them to 1 results in $\sigma(S^0) \sim 2.7,7.5$~pb  for $M_S= 150, 500$~GeV respectively at LHC13 \footnote{We thank Alper Hayreter for providing these numbers.}. The main difference being that one of these values is above the $t\bar{t}$ threshold. 

The SM one-loop result, $m_ta_t^g=-0.014$, is an order of magnitude larger than typical corrections predicted by this model, and is comparable to corrections in 2HDM or models with extra dimensions as computed in Ref.~\cite{Martinez:2007qf}.  The corrections calculated here and listed in Table~\ref{t:pre} are labelled `typical' as they can be pushed up by an order of magnitude by allowing couplings such as $\eta_U$ to be as large as their unitarity upper bound and/or considering lighter scalars which at the moment are not ruled out. 

Existing models with vector like multiplets \cite{Ibrahim:2011im} predict $d_t^g$ as large as 0.003   a bit above our typical numbers. The range for  $d_t^g$ predicted in this model is below the reach of near future experiments, as are the b-quark couplings $d_b^g$ and $a_b^g$.

The typical corrections to $f_T^R$ in this model as shown in Table~\ref{t:pre} are  about a factor of three smaller than the SM one-loop value $f_T^R=-(7.17+1.23i)\times 10^{-3}$ as calculated in Ref.~\cite{GonzalezSprinberg:2011kx}. A major difference from results obtained in 2HDM \cite{Duarte:2013zfa} is that in the MW model the correction to $f_T^R$ is much larger than that to $f_T^L$ except for regions of parameter space where $\eta_U < \eta_D m_b/m_t$.

The typical values shown in Table~\ref{t:pre} are below the potential constraints illustrated in Table~\ref{t:results}. But it is clear that as the experimental constraints approach the numbers in the Table, they will begin to limit the parameter space of the MW model beyond perturbative unitarity.

\begin{acknowledgments}

R.M. was supported by El Patrimonio Aut\'{o}nomo Fondo Nacional de Financiamiento para la Ciencia, la Tecnolog\'{i}a y la Innovaci\'{o}n Francisco Jos\'{e} de Caldas programme of COLCIENCIAS in Colombia.  G.V. thanks the Departamento de F\'{i}sica, Universidad Nacional de Colombia for their hospitality while part of this work was completed.

\end{acknowledgments}

\appendix

\section{Feynman parameter integrals}

All the loop diagrams appearing in the EDM, MDM, CEDM and CMDM calculations presented in this paper can be written in terms of the one parameter integrals
\begin{eqnarray}
F_{n,m}(r)&=&\int_0^1\frac{x^n(1-x)^m}{1-x+r^2x^2}dx\nonumber \\
G_{n,m}(r_1,r_2)&=&\int_0^1\frac{x^n(1-x)^m}{(1-x)(1-r_1^2x)+r_2^2x}dx  
\end{eqnarray}

Although these integrals can be evaluated analytically, for the purposes of this paper we choose to evaluate numerically the combinations that appear in the final results. For convenience we present here the limiting values for small values of $r$, corresponding to large values of $M_S$.
\begin{eqnarray}
 F_{0,0}(r\to 0)&=&-4\log(r)-3         \nonumber \\
 F_{2,0}(r\to 0)&=&-2\log(r)-\frac{11}{6}       \nonumber \\
F_{3,0}(r\to 0)&=&-2\log(r)-\frac{3}{2}        \nonumber \\
  F_{1,1}(r\to 0)&=&  \frac{1}{2} \nonumber \\
    F_{1,2}(r\to 0)&=& \frac{1}{6}  \nonumber \\
  F_{2,1}(r\to 0)&=&\frac{1}{3} \nonumber \\
G_{n,m}(r_1\to 0,r_2\to 0) &=& F_{n,m}(r\to 0)
 \end{eqnarray}
 For the calculation of $f_T^R$ we need the two parameter Feynman integral
\begin{eqnarray}
F_c(r_t,r_W)=\int_0^1dx\int_0^{1-x}dy\frac{2y}{1-y+r_t^2y^2-r_W^2r_t^2x(1-x)+r_t^2(1+r_W^2)xy}
\label{frtrw}
\end{eqnarray}
Some manipulations have been carried out with the help of FeynCalc \cite{Shtabovenko:2016sxi}.

\end{document}